\documentclass[fleqn,10pt]{wlscirep}
\usepackage{epsfig}
\usepackage{subfigure}
\usepackage{amssymb,amsmath}
\usepackage{graphicx}
\usepackage{epstopdf}
\usepackage{epsfig}
\usepackage{color}
\usepackage{amsfonts}
\usepackage{amscd}
\usepackage{hyperref}
\usepackage{bbm}
\usepackage{mathrsfs}
\title{Dark state with counter-rotating dissipative channels}

\author[1,2]{Zheng-Yang Zhou}

\author[1,2]{Mi Chen}
\author[3,4]{Lian-Ao Wu}

\author[2,5]{Ting Yu}

\author[2,*]{J. Q. You}

\affil[1]{Department of Physics, Fudan University, Shanghai 200433, China}
\affil[2]{Beijing Computational Science Research Center, Beijing
100094, China}
\affil[3]{Ikerbasque, Basque Foundation for Science, 48011 Bilbao, Spain}
\affil[4]{Department of Theoretical Physics and History of Science,The Basque Country University (EHU/UPV),
P. O. Box 644, 48080 Bilbao, Spain}
\affil[5]{Center for Controlled Quantum Systems and Department of Physics and Engineering Physics,
Stevens Institute of Technology, Hoboken, New Jersey 07030, USA}
\affil[*] {Correspondence and requests for materials should be addressed to J.Q.Y. (email: jqyou@csrc.ac.cn)}

\begin{abstract}
Dark state as a consequence of interference between different quantum states has great importance in the fields of chip-scale atomic clock and quantum information. For the $\Lambda$-type three-level system, this dark state is generally regarded as being dissipation-free because it is a superposition of two lowest states without dipole transition between them. However, previous studies are based on the rotating-wave approximation (RWA) by neglecting the counter-rotating terms in the system-environment interaction. In this work, we study non-Markovian quantum dynamics of the dark state in a $\Lambda$-type three-level system coupled to two bosonic baths and reveal the effect of counter-rotating terms on the dark state.
In contrast to the dark state within the RWA, leakage of the dark state occurs even at zero temperature, as a result of these counter-rotating terms. Also, we present a method to restore the quantum coherence of the dark state by applying a leakage elimination operator to the system.
\end{abstract}
\begin{document}

\flushbottom
\maketitle
%

\thispagestyle{empty}

\section*{Introduction}

Electromagnetically induced transparency discovered in quantum optics has long been an important effect in physics (see, e.g., \cite{eit} for a review). This phenomenon of absorption cancelation is interpreted as the appearance of dark state or coherent population trapping. In addition to the atomic systems, dark state has also been observed in a number of solid-state systems including quantum dots~\cite{qdds,qdds2}, nitrogen-vacancy center~\cite{nvc} and silicon-vacancy center in diamond~\cite{sv1,sv2}. In fact, dark state can have different applications in physics. The atomic clocks based on coherent population trapping~\cite{ac1,ac2,ac3,ac4} make the high-precision time estimation possible using the chip-scale and low-power devices. The state transfer can be done with the adiabatic passage of the dark state~\cite{ap1,ap2,ap3}. Operations on the quantum states like squeezing~\cite{sq} or decay suppression~\cite{ds1,ds2} can also be conducted with the help of dark state. Moreover, dark state can have important applications to the slow light~\cite{slight} and photocell~\cite{pcell}.

Here we consider the $\Lambda$-type three-level system where dipole transition between the lowest two states is forbidden. Within the framework of rotating-wave approximation (RWA) for the interaction between the system and the environment, the dark state is dissipation-free at a low enough temperature. For instance, the dark state is not influenced by the spontaneous emission~\cite{eit}. Studies in two-level systems have shown that the counter-rotating terms can change the ground state~\cite{twolevel1,twolevel2,twolevel3}, but there were few studies regarding the influence of the counter-rotating terms on the dark state~\cite{cr}. When the coupling between the system and the environment becomes strong, the counter-rotating terms cannot be neglected. Thus, interesting phenomena with the quantum dynamics of the dark state are expected even at zero temperature, because now the ground state within the framework of RWA is no longer the ground state of the system when including the counter-rotating terms.

In this article, we study the quantum dynamics of the dark state beyond the RWA, where the $\Lambda$-type three-level system couples to two bosonic baths at zero temperature and the couplings between the system and the two baths contain both rotating and counter-rotating terms. We derive a non-Markovoian quantum Bloch equation for the dark state using a quantum Langevin approach. In contrast to the dark state within the RWA, leakage of the dark state occurs due to the counter-rotating terms in the system-bath interaction, revealing the breakdown of the dissipation-free dark state at the zero temperature. To suppress the leakage, we apply a leakage elimination operator to the system, which plays the role of keeping the upper level of the $\Lambda$-type three-level system empty. Indeed, the leakage of the dark state can be much reduced when applying the elimination operator, as shown in our numerical results. This study provides a method to restore the quantum coherence of the dark state with counter-rotating dissipative channels.

\section*{Results}

\section{The model Hamiltonian}
\begin{figure}
\center
\includegraphics[width=3.4in]{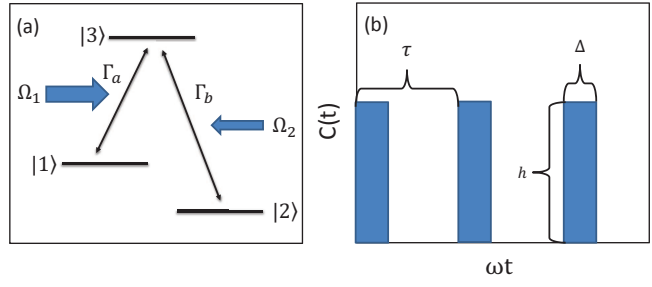}
\caption{(a) A $\Lambda$-type three-level system driven by two fields with frequencies $\omega_a$ and $\omega_b$, respectively. The field on the left (right), which drives the transition between $|1\rangle$ ($|2\rangle$) and $|3\rangle$, has a coupling strength $\Omega_1$ ($\Omega_2$) with this transition. The three-level system is also coupled to two bosonic baths, with the coupling strengths characterized by $\Gamma_a$ and $\Gamma_b$.
(b) Schematic illustration of the applied control pulses, where $\tau$ is the period of the pulses, $\Delta$ is the duration of each pulse, and $h$ is the strength of the pulse.}
\label{tl}
\end{figure}

We study a $\Lambda$-type three-level system driven by two pump fields [see Fig.~\ref{tl}(a)]. The Hamiltonian of this three-level system is given by (setting $\rm\hbar=1$)
\begin{eqnarray}
H_{\rm sys}&=&\omega_1|1\rangle\langle1|+\omega_2|2\rangle\langle2|+\omega_3|3\rangle\langle3|,\nonumber\\
&&+(\Omega_1 e^{-i\omega_{a}t}|3\rangle\langle1|+\Omega_2 e^{-i\omega_{b}t}|3\rangle\langle2|+\rm{H.c.}),~~~
\label{htl}
\end{eqnarray}
where $\omega_1$, $\omega_2$, and $\omega_3$ are the three energy levels of the system and the frequencies of the two pumping fields are $\omega_{a}$ and $\omega_{b}$, respectively, with $\Omega_1$ and $\Omega_2$ characterizing their coupling strengths to the three-level system. To focus on the effect of the counter-rotating dissipative channels, we take $\Omega_1$ and $\Omega_2$ to be independent of time. Also, the effects like dephasing channels or non-adiabatic transitions are not included here because they were well studied~\cite{ap3}. Here we consider the resonant \textcolor{red}{pumping} case with $\omega_{a}=\omega_3-\omega_1$ and $\omega_{b}=\omega_3-\omega_2$.
This $\Lambda$-type three-level system has a dark state \cite{eit},
\begin{equation}
|D(t)\rangle=\frac{\Omega_2}{\Omega}e^{-i\omega_1t}|1\rangle-\frac{\Omega_1}{\Omega}e^{-i\omega_2t}|2\rangle,
\label{dark}
\end{equation}
with $\Omega=\sqrt{|\Omega_1|^2+|\Omega_2|^2}$, which is a solution of the Schr{\" o}dinger equation,
$\frac{\partial}{\partial t}|D(t)\rangle=-iH_{\rm sys}|D(t)\rangle$.
When the three-level system is in this state, it remains in the subspace spanned by $\{|1\rangle, |2\rangle\}$, even in the presence of the two pumping fields.

To study the effect of the environments on the dark state, we use two independent bosonic baths modeled by
$H_{\rm bath}=\sum_k\omega_{a,k}a^{\dag}_ka_k+\sum_k\omega_{b,k}b^{\dag}_kb_k$. The interaction between the system and the two baths is
\begin{eqnarray}
H_{\rm int}&=&(|1\rangle\langle3|+|3\rangle\langle1|)\sum_k(g_{a,k}a_k+g_{a,k}^*a_k^{\dag})\nonumber\\
&&+(|2\rangle\langle3|+|3\rangle\langle2|)\sum_k(g_{b,k}b_k+g_{b,k}^*b_k^{\dag}),
\label{eq-int}
\end{eqnarray}
where $a_k$ and $b_k$ are annihilation operators of the bosonic modes of the two baths. Note that both rotating and counter-rotating terms are included in this interaction Hamiltonian. The dipole transition between $|1\rangle$ and $|2\rangle$ is forbidden in the considered $\Lambda$-type three-level system, so we omit this channel here. Under the RWA (i.e., only the rotating terms are considered), Eq.~(\ref{eq-int}) is reduced to
\begin{equation}
H_{\rm RWA}=\sum_k(g_{a,k}a_k|3\rangle\langle1|+g_{b,k}b_k|3\rangle\langle2|)+{\rm H.c.}.
\end{equation}
Let the two bosonic baths be in the vacuum state $|0\rangle\equiv|0\rangle_a\otimes|0\rangle_b$, which corresponds to the zero temperature for each bath. It is easy to check that $(H_{\rm RWA}+H_{\rm bath})|D(t)\rangle\otimes|0\rangle=0$. Thus,
\begin{equation}
\frac{\partial}{\partial t}|D(t)\rangle\otimes|0\rangle=-i(H_{\rm sys}+H_{\rm RWA}+H_{\rm bath})|D(t)\rangle\otimes|0\rangle,
\label{RWA-dark}
\end{equation}
i.e., the dark state can persist even when the three-level system couples to the two baths. However, the counter-rotating terms in Eq.~(\ref{eq-int}) can make the dark state transition to the state $|3\rangle$, so $|D(t)\rangle\otimes|0\rangle$ is not a solution of the Schr{\"o}dinger equation,
$\frac{\partial}{\partial t}|\Psi\rangle=-iH_{\rm tot}|\Psi\rangle$,
where $H_{\rm tot}=H_{\rm sys}+H_{\rm int}+H_{\rm bath}$
is the total Hamiltonian of the system.

Below we show how the two baths affect the quantum coherence of the dark state when the interaction Hamiltonian is $H_{\rm int}$, so as to reveal the effect of the counter-rotating terms in the interaction Hamiltonian. Then, we present a method to restore the quantum coherence of the dark state by applying a leakage elimination operator to the system.

\section{Non-Markovian quantum Bloch equation}
The reduced density operator of the system can be written as
\begin{equation}
\rho_{\rm sys}(t)=\sum_{i,j}\rho^{\rm (sys)}_{ij}(t)|i\rangle\langle j|,
\end{equation}
where $i,j=1,2,3$, and the reduced density matrix elements are given by
\begin{equation}
\rho^{\rm (sys)}_{ij}(t)=\langle i|{\rm Tr}_{\rm env}\rho(t)|j\rangle.
\end{equation}
Here $\rho(t)$ is the density operator of the total system and ${\rm Tr}_{\rm env}$ denotes the trace over the degrees of freedom of the environments. These reduced density matrix elements can be rewritten as
\begin{eqnarray}
\rho^{\rm (sys)}_{ij}(t)&=&\sum_{n=1}^3\langle i|n\rangle\langle n|{\rm Tr}_{\rm env}\rho(t)|j\rangle \nonumber\\
&=&\sum_{n=1}^3\langle n|{\rm Tr}_{\rm env}\rho(t)|j\rangle\langle i|n\rangle \nonumber\\
&=&{\rm Tr}\left[\rho(t)|j\rangle\langle i|\right],
\end{eqnarray}
where ${\rm Tr}$ denotes the trace over the degrees of freedom of the total system. Thus, $\rho^{\rm (sys)}_{ij}(t)$ is just the expectation value of the system operator $|j\rangle\langle i|$ and it can also be written, in the Heisenberg picture, as
\begin{eqnarray}
\rho^{\rm (sys)}_{ij}(t)&=&\langle\Psi(t)|j\rangle\langle i|\Psi(t)\rangle \nonumber\\
&=&\langle\Psi_0|j\rangle\langle{i}|(t)|\Psi_0\rangle,
\end{eqnarray}
with
\begin{equation}
|j\rangle\langle{i}|(t)\equiv U^{\dag}(t)|j\rangle\langle{i}|U(t),
\end{equation}
where $|j\rangle\langle i|(t)$ is a system operator represented in the Heisenberg picture and $|j\rangle\langle i|$ is this operator at the initial time $t=0$, while $U(t)$ is the evolution operator,
\begin{equation}
U(t)=Te^{-i\int_0^tH_{\rm tot}(s)ds},
\label{evolution}
\end{equation}
with $T$ being the time-ordering operator. Here we study the case with the initial state of the total system given $|\Psi_0\rangle\equiv|D(0)\rangle\otimes|0\rangle$.

To conveniently see the dynamical behavior of the system from non-Markovian to Markovian, we choose the correlation functions $\alpha_i(t,s)\equiv\sum_k|g_{i,k}|^2e^{-i\omega_k(t-s)}$ of the two baths as the typical Ornstein-Uhlenbeck correlation functions, $\alpha_i(t,s)=\frac{\Gamma_i\gamma_i}{2}e^{-\gamma_i|t-s|}$, where $i=a,b$. The non-Markovian to Markovian transition can be demonstrated by tuning the parameters $\gamma_i$, i.e., the inverse of the correlation times of the two baths. The coupling strength between the system and the $i$th bath is given by $\Gamma_i$ which corresponds to the decay rate under Markovian approximation~\cite{qsd1}.
Using the Heisenberg equation, we can derive the following non-Markovian quantum Bloch equation for the expectation values of the system operators (see method):
\begin{equation}
\frac{\partial}{\partial t}\mathcal{A}(t)=\mathcal{H}(t)\mathcal{A}(t)+\mathcal{L}_a\mathcal{A}^{(1,0)}(t)
+\mathcal{L}_b\mathcal{A}^{(0,1)}(t),
\label{qbe}
\end{equation}
where $\mathcal{A}(t)\equiv(\mathcal{A}_{ij}(t))$, with $i,j=1,2,3$, and $\mathcal{A}_{ij}(t)=\langle\Psi_0|i\rangle\langle j|(t)|\Psi_0\rangle\equiv\rho^{\rm (sys)}_{ij}(t)$. In Eq.~(\ref{qbe}), $\mathcal{H}(t)$, $\mathcal{L}_a$ and $\mathcal{L}_b$ are given by
\begin{equation}
\mathcal{H}(t)=\left(\begin{array}{ccccccccc}
               0&0&0&i\Omega_1(t)&0&0&-i\Omega_1^*(t)&0&0\\
               0&0&0&0&i\Omega_2(t)&0&0&-i\Omega_2^*(t)&0\\
               0&0&0&-i\Omega_1(t)&-i\Omega_2(t)&0&i\Omega_1^*(t)&i\Omega_2^*(t)&0\\
               i\Omega_1^*(t)&0&-i\Omega_1^*(t)&i\Delta_{31}&0&i\Omega_2^*(t)&0&0&0\\
               0&i\Omega_2^*(t)&-i\Omega_2^*(t)&0&i\Delta_{32}&0&0&0&i\Omega_1^*(t)\\
               0&0&0&i\Omega_2(t)&0&i\Delta_{21}&0&-i\Omega_1^*(t)&0\\
               -i\Omega_1(t)&0&i\Omega_1(t)&0&0&0&-i\Delta_{31}&0&-i\Omega_2(t)\\
               0&-i\Omega_2(t)&i\Omega_2(t)&0&0&-i\Omega_1(t)&0&-i\Delta_{32}&0\\
               0&0&0&0&i\Omega_1(t)&0&-i\Omega_2^*(t)&0&-i\Delta_{21}\\
               \end{array}\right),
\label{matrixH}
\end{equation}
\begin{equation}
\mathcal{L}_a=\left(\begin{array}{ccccccccc}
               0&0&0&i&0&0&-i&0&0\\
               0&0&0&0&0&0&0&0&0\\
               0&0&0&-i&0&0&i&0&0\\
               i&0&-i&0&0&0&0&0&0\\
               0&0&0&0&0&0&0&0&i\\
               0&0&0&0&0&0&0&-i&0\\
               -i&0&i&0&0&0&0&0&0\\
               0&0&0&0&0&-i&0&0&0\\
               0&0&0&0&i&0&0&0&0\\
               \end{array}\right),~~~~~~
\mathcal{L}_b=\left(\begin{array}{ccccccccc}
               0&0&0&0&0&0&0&0&0\\
               0&0&0&0&i&0&0&-i&0\\
               0&0&0&0&-i&0&0&i&0\\
               0&0&0&0&0&i&0&0&0\\
               0&i&-i&0&0&0&0&0&0\\
               0&0&0&i&0&0&0&0&0\\
               0&0&0&0&0&0&0&0&-i\\
               0&-i&i&0&0&0&0&0&0\\
               0&0&0&0&0&0&-i&0&0\\
               \end{array}\right).
\label{matrixL}
\end{equation}
where $\Delta_{ij}=\omega_i-\omega_j\ (i,j=1,2,3)$ and $\Omega_i(t)=\Omega_ie^{-i\Delta_{3i}t}\ (i=1,2)$.
The non-Markovianity of the quantum dynamics of the three-level system is reflected in both $\mathcal{A}^{(1,0)}(t)$ and $\mathcal{A}^{(0,1)}(t)$, which are solved via the hierarchical equation
\begin{eqnarray}
\frac{\partial}{\partial t}\mathcal{A}^{(m,n)}(t)
                        &=&-(m\gamma_a+n\gamma_b)\mathcal{A}^{(m,n)}(t)+\mathcal{H}(t)\mathcal{A}^{(m,n)}(t)\nonumber\\
                        &&+m\frac{\Gamma_a\gamma_a}{2}\mathcal{L}_a\mathcal{A}^{(m-1,n)}(t)
                        +\mathcal{L}_a\mathcal{A}^{(m+1,n)}(t)\nonumber\\
                         &&+n\frac{\Gamma_b\gamma_b}{2}\mathcal{L}_b\mathcal{A}^{(m,n-1)}(t)
                         +\mathcal{L}_b\mathcal{A}^{(m,n+1)}(t),\nonumber\\
\label{be2}
\end{eqnarray}
where $\mathcal{A}^{(m,n)}(t)=0$ if $m$ or $n<0$, $\mathcal{A}^{(0,0)}(t)\equiv\mathcal{A}(t)$, and the initial condition is $\mathcal{A}^{(m,n)}(0)=0$ for $m$ or $n\ne 0$.

With the reduced density operator $\rho_{\rm{sys}}(t)$ obtained by choosing the initial state of the total system as $|\Psi_0\rangle\equiv|D(0)\rangle\otimes|0\rangle$, the fidelity of the dark state of the three-level system can be written as
\begin{equation}
\mathcal{F}(t)\equiv\sqrt{\langle{D(t)}|\rho_{\rm{sys}}(t)|D(t)\rangle}.
\end{equation}
This quantity can be used to characterize the leakage of the dark state to other levels.

\section{Breakdown of the dark state}
\begin{figure}
\center
\includegraphics[width=4.5in]{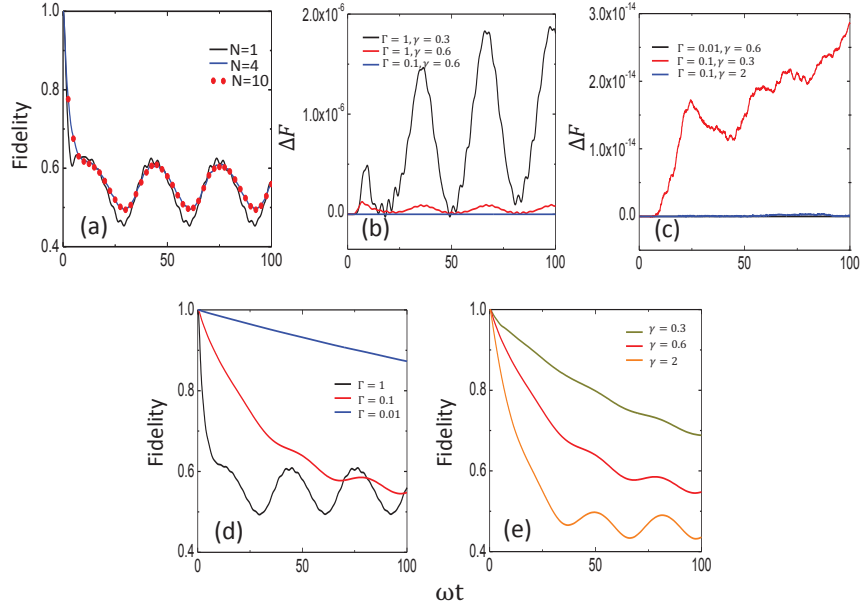}
\caption{Fidelity evolution of the dark state. (a)~Fidelity of the dark state calculated at different order $\mathcal{N}$ of the hierarchical equation in Eq.~(\ref{be2}). The parameters of the two baths are $\Gamma=1$ and $\gamma=0.6$.
(b)~Fidelity difference $\Delta F\equiv F_{10}-F_{20}$, where $F_{10}$ ($F_{20}$) is the fidelity of the dark state calculated with the hierarchical equation up to the order of $\mathcal{N}=10$ (20). (c)~Fidelity of the dark state calculated at different values of $\Gamma$, where $\gamma=0.6$ and $\mathcal{N}=10$. (d)~Fidelity of the dark state calculated at different values of $\gamma$, where $\Gamma=0.1$ and $\mathcal{N}=10$.
In this figure and the following one, the parameters of the three-level system are $\omega_1=-1$, $\omega_2=-1.2$, $\omega_3=0$, $\Omega_1=0.5$, and $\Omega_2=0.2$. }
\label{fig2}
\end{figure}

As shown in Eq.~(\ref{RWA-dark}), for the interaction Hamiltonian $H_{\rm RWA}$ with only the rotating terms, the dark state persists when the three-level system couples to two zero-temperature baths. Thus, $\mathcal{F}(t)=1$ in this case. Below we demonstrate the dynamical evolution of the fidelity $\mathcal{F}(t)$ of the dark state when the counter-rotating terms are included in the interaction Hamiltonian. For simplicity, we choose the same parameters for the two baths, i.e., $\alpha_a(t,s)=\alpha_b(t,s)=\frac{\Gamma\gamma}{2}e^{-\gamma|t-s|}$. We also choose the level of state $|3\rangle$ to be the zero point of the energy. The energy difference between $|1\rangle$ and $|3\rangle$ is taken as $\omega_3-\omega_1=\omega=1$. Other parameters with the frequency unit are expressed as the ratio to $\omega$. To numerically calculate the quantum dynamics of the system, we need to truncate Eq.~(\ref{be2}) at a given hierarchical order $\mathcal{N}$ so that $\mathcal{A}^{(m,n)}(t=0)$ for $m+n>\mathcal{N}$. As shown in Fig.~\ref{fig2}(a), the numerical results at $\mathcal{N}=4$ (blue curve) are very close to the result at $\mathcal{N}=10$ (read dots), indicating that the results converge rapidly with the hierarchical order $\mathcal{N}$. Note that in this case, we choose $\Gamma=1$, corresponding to a strong coupling between the system and the two baths (the corresponding Markovian decay rate is now comparable to the system frequencies). When $\Gamma<1$, more accurate results are obtained at the same hierarchical order due to the even faster convergence of the numerical results in the weak coupling regime between the system and the baths.
Thus, the truncation of Eq.~(\ref{be2}) at $\mathcal{N}=4$ can already give reliable results for the model that we study here. However, we choose $\mathcal{N}=10$ in our following calculations to have the obtained results more accurate. Also, Fig.~\ref{fig2}(b) shows the difference between the results of $\mathcal{N}=10$ and $\mathcal{N}=20$. It is found that the difference is below $10^{-5}$, further indicating that the truncation is reliable.

In Fig.~\ref{fig2}(c), we show the fidelity evolution of the dark state by varying the coupling strength $\Gamma$ between the system and the two baths. When $\Gamma=1$, the fidelity of the dark state decays fast and then quickly reaches the stationary oscillations in this strong system-bath coupling regime. These stationary oscillations correspond to a dynamic equilibrium of the dark state under the combined actions of the drive fields and the baths. However, when decreasing the coupling strength $\Gamma$, the fidelity of the dark state decays slowly and takes a long time to reach the stationary oscillations. Thus, with a given correlation time (inverse of $\gamma$) of the two baths, the dark state has a slow leakage with time when the system-bath coupling is weak.
In Fig.~\ref{fig2}(d), we also show the effect of the environment correlation time (i.e., $\gamma$) on the fidelity evolution of the dark state. For the non-Markovian environment, which has a longer correlation time (i.e., smaller $\gamma$), the fidelity of the dark state decays slowly with the evolution $t$, in comparison with the Markovian environment with a shorter correlation time (i.e., larger $\gamma$). This reveals that the dark state leaks to other levels more slowly when it is coupled to a non-Markovian environment. Moreover, similar to Fig.~\ref{fig2}(c), the fidelity of the dark state exhibits stationary oscillations at longer times even in the Markovian limit (large $\gamma$) of the baths. These oscillations are also due to the persistent drive applied to the system.

From our results above, we can conclude that the dark state is unstable under the influence of the counter-rotating terms even when the environments are at zero temperature. Thus, owing to the counter-rotating terms in the interaction Hamiltonian, the environment-induced effect on the dynamical evolution of the dark state cannot be eliminated by simply lowering the temperature of the environments.

\section{Leakage reduction of the dark state}
To reduce the leakage of the dark state, we introduce a leakage elimination operator~\cite{elo1,elo2,elo3,elo4,elo5,elo6,elo7}. When this leakage elimination operator is added, the total Hamiltonian of the open system becomes
\begin{equation}
H_{\rm{tot}}=H_{\rm{sys}}+H_{\rm{bath}}+H_{\rm{int}}+R(t),
\end{equation}
where
\begin{equation}
R(t)=-c(t)(|1\rangle\langle1|+|2\rangle\langle2|)
 \label{LEO}
\end{equation}
is the leakage elimination operator that suppresses leakage from the system to the environment. In numerical calculations, we use the rectangular control pulse $c(t)$ as an example, which has a period of $\tau$ [see Fig.~\ref{tl}(b)]. Within the time intervals $l\tau\leq t<l\tau+\Delta$, where $l$ is a positive integer to identify the pulse is in the $l$th period, the control pulse is switched on and has an intensity of $h$. For other times, the control pulse is turned off.

\begin{figure}
\center
\includegraphics[width=4.5in]{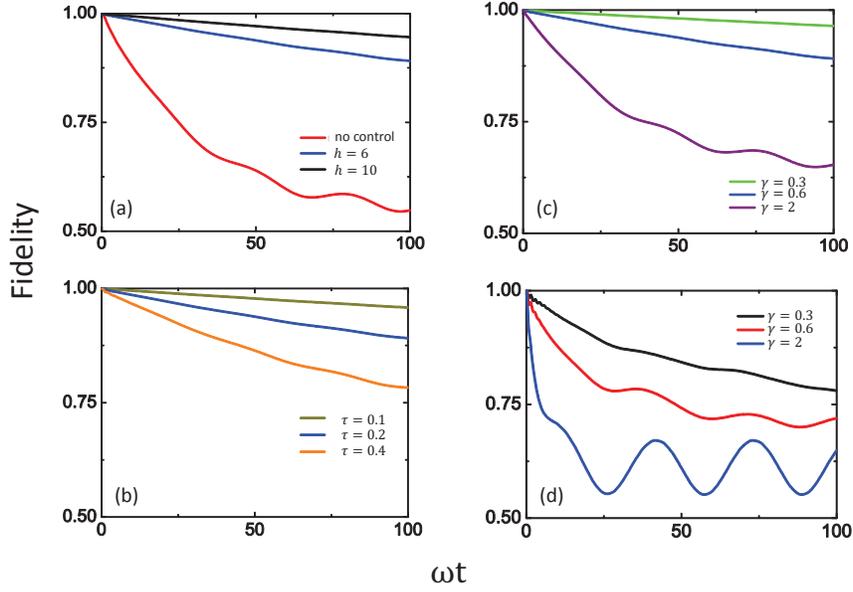}
\caption{Fidelity evolution of the dark state in the presence of the control pulses. (a)~Fidelity of the dark state when varying the strength of the pulse, where the period of the pulses is $\tau=0.2$ and the duration of each pulse is $\Delta=0.1$. The parameters of the two baths are $\Gamma=0.1$ and $\gamma=0.6$. (b)~Fidelity of the dark state when varying the period of the pulses, where the strength of the pulse is $h=0.6/\Delta$ and the duration is chosen to be $\Delta=0.5\tau$. The parameters of the two baths are also $\Gamma=0.1$ and $\gamma=0.6$. (c)~Fidelity of the dark state when varying the inverse of the bath correlation time $\gamma$, where the strength of the pulse is $h=6$, the period of the pulses is $\tau=0.2$, and the duration of each pulse is $\Delta=0.1$. The coupling strength of each bath is $\Gamma=0.1$. (d)~Fidelity of the dark state when varying the inverse of the bath correlation time $\gamma$, where the coupling strength of each bath to the system is $\Gamma=1$. Other parameters are the same as in (c).}
\label{ll}
\end{figure}

From Fig.~\ref{ll}(a), it can be seen that when the control pulse is applied, the fidelity of the dark state (black and blue curves) decreases much slowly with the evolution time, in sharp contrast with the fidelity without the control pulse  (red curve). Thus, the leakage elimination operator works quite effectively in reducing the leakage of the dark state. Also, the black curve has a higher fidelity than the blue one, implying that a higher pulse intensity yields better control effect for suppressing the state leakage. Figure~\ref{ll}(b) presents the effect of the period of the control pulse on the leakage elimination. It is clear that the fidelity of dark state increases when decreasing the period of the control pulse. Therefore, both higher intensity and frequency of the control pulse can strengthen the effect of the leakage elimination on the dark state. In Figs.~\ref{ll}(c) and \ref{ll}(d), we further show the effect of the inverse of the correlation time $\gamma$. Compared with Fig.~\ref{fig2}(c), it can be seen that the leakage of the dark state is much eliminated for very non-Markovian baths with small values of $\gamma$. This is the distinct advantage of the non-Markovian baths.

Finally, we discuss more about the leakage elimination operator. Instead of using the eigenstates $|1\rangle$, $|2\rangle$ and $|3\rangle$, we can also use the dark state $|D(t)\rangle$, bright state $|B(t)\rangle$ and eigenstate $|3\rangle$ as the basis states of the Hilbert space of the three-level system, where the dark state is given in Eq.~(\ref{dark}) and the bright state is
\begin{equation}
|B(t)\rangle=\frac{\Omega_1}{\Omega}e^{-i\omega_1t}|1\rangle+\frac{\Omega_2}{\Omega}e^{-i\omega_2t}|2\rangle.
\end{equation}
With this set of basis states, we can rewrite the interaction Hamiltonian in Eq.~(\ref{eq-int}) as
\begin{eqnarray}
H_{\rm int}\!&\!=\!&\!\frac{\Omega_2^*}{\Omega}e^{i\omega_1t}|D(t)\rangle\langle3|\sum_k(g_{a,k}a_k+g_{a,k}^*a_k^{\dag})\nonumber\\
     &&\!-\frac{\Omega_1^*}{\Omega}e^{i\omega_2t}|D(t)\rangle\langle3|\sum_k(g_{b,k}b_k+g_{a,k}^*b_k^{\dag})\nonumber\\
     &&\!+\frac{\Omega_1^*}{\Omega}e^{i\omega_1t}|B(t)\rangle\langle3|\sum_k(g_{a,k}a_k+g_{a,k}^*a_k^{\dag})\nonumber\\
     &&\!+\frac{\Omega_2^*}{\Omega}e^{i\omega_2t}|B(t)\rangle\langle3|\sum_k(g_{b,k}b_k+g_{a,k}^*b_k^{\dag})+\mathrm{H.c.}.
     ~~~~~~~
\label{int}
\end{eqnarray}
From Eq.(\ref{int}), it follows that there is no direct leakage from the dark state to the bright state, so the leakage of the dark state occurs only via the excited state $|3\rangle$. If the state $|3\rangle$ is kept unoccupied, the leakage of the dark state can be prevented. That is why we can use the leakage elimination operator in Eq.~(\ref{LEO}) to restore the quantum coherence of the dark state. The leakage elimination operator in Eq.~(\ref{LEO}) also have the effect of counterdiabatic driving~\cite{cadia1,cadia2,cadia3,cadia4}. Because it can effectively enlarge the energy difference of the system, the adiabatic condition can be better fulfilled. Here we have proved the equivalence of the leakage elimination operator for $|3\rangle$ to the one for $|D(t)\rangle$ in the case of dissipative channels, but the equivalence about the counterdiabatic effects is not clear.

\section*{Discussion}
We have studied the non-Markovian quantum dynamics of the dark state in a $\Lambda$-type three-level system coupled to two bosonic baths and revealed the effect of the counter-rotating terms in the system-bath coupling on the dark state.
 Due to these counter-rotating terms, the dark state leaks to other states even at zero temperature, in sharp contrast to the dark state within the RWA. Thus, whether the dark state is really dark or not depends on the validity of the RWA in the considered system. To restore the quantum coherence of the dark state, we propose to apply a leakage elimination operator to the system. Our numerical results indicate that the leakage of the dark state can indeed be greatly suppressed  with the help of this leakage elimination operator.

Our study reveals a possible mechanism for the dark state to leak and a way to fight against it. This may improve the precision of the experiments related to the dark state.  While the studies of quantum dynamics beyond the RWA mainly focus on the two-level systems, our work provides insights into the dynamics of the three-level system beyond the RWA.
Also, people can compare the leakage indicated in our article with the leakage induced by other mechanisms (e.g., dephasing channel, finite temperature effect or non-adiabatic transition) to see whether the effect of counter-rotating terms can be neglected in their cases.

\section*{Methods}
From the total Hamiltonian $H_{\mathrm{tot}}$ given in Sec.~II, it follows that the Heisenberg equations for the field operators $a_k$ and $b_k$ are given by
\begin{eqnarray}
\frac{d}{dt}a_k(t)&=&-i\omega_ka_{k}(t)-ig_{a,k}^*(|1\rangle\langle3|(t)+|3\rangle\langle1|(t)),\nonumber\\
\frac{d}{dt}b_k(t)&=&-i\omega_kb_k(t)-i{g}_{b,k}^*(|2\rangle\langle3|(t)+|3\rangle\langle2|(t)),
\label{ebo}
\end{eqnarray}
were $a_k(t)\equiv U^{\dag}(t)a_kU(t)$ and $b_k(t)\equiv U^{\dag}(t)b_kU(t)$, with $U(t)$ given in Eq.~(\ref{evolution}). Equation~(\ref{ebo}) can be formally solved as
\begin{eqnarray}
a_k(t)&=&e^{-i\omega_kt}a_k-ig_{a,k}^*\int_0^tdse^{-i\omega_k(t-s)}
\left[|1\rangle\langle3|(s)+|3\rangle\langle1|(s)\right]\nonumber,\\
b_k(t)&=&e^{-i\omega_kt}b_k-i{g}_{b,k}^*\int_0^tdse^{-i\omega_k(t-s)}
\left[|2\rangle\langle3|(s)+|3\rangle\langle2|(s)\right].\nonumber\\
\label{sbo}
\end{eqnarray}
Similar to Eq.~(\ref{ebo}), we can also derive the Heisenberg equations for the system operators $|i\rangle\langle j|$, with $i,j=1,2,3$. Then, substituting Eq.~(\ref{sbo}) into the Heisenberg equations for the system operators, we have
\begin{eqnarray}
\frac{d}{dt}|1\rangle\langle1|(t)&=&i\Omega_1e^{-i(\omega_3-\omega_1)t}|3\rangle\langle1|(t)-i\Omega_1^*e^{i(\omega_3-\omega_1)t}|1\rangle\langle3|(t)+i(|3\rangle\langle1|(t)-|3\rangle\langle1|(t))(\xi_a(t)+\xi_a^{\dag}(t)),\nonumber\\
\frac{d}{dt}|2\rangle\langle2|(t)&=&i\Omega_2e^{-i(\omega_3-\omega_2)t}|3\rangle\langle2|(t)-i\Omega_2^*e^{i(\omega_3-\omega_2)t}|2\rangle\langle3|(t)+i(|3\rangle\langle2|(t)-|2\rangle\langle3|(t))(\xi_b(t)+\xi_b^{\dag}(t)),\nonumber\\
\frac{d}{dt}|3\rangle\langle3|(t)&=&i\Omega_1^*e^{i(\omega_3-\omega_1)t}|1\rangle\langle3|(t)-i\Omega_1e^{-i(\omega_3-\omega_1)t}|3\rangle\langle1|(t)+i\Omega_2^*e^{i(\omega_3-\omega_2)t}|2\rangle\langle3|(t)-i\Omega_2e^{-i(\omega_3-\omega_2)t}|3\rangle\langle2|(t)\nonumber\\
                               &&+i(|1\rangle\langle3|(t)-|3\rangle\langle1|(t))(\xi_a(t)+\xi_a^{\dag}(t))+i(|2\rangle\langle3|(t)-|3\rangle\langle2|(t))(\xi_b(t)+\xi_b^{\dag}(t)),\nonumber\\
\frac{d}{dt}|3\rangle\langle1|(t)&=&i(\omega_3-\omega_1)|3\rangle\langle1|(t)+i\Omega_1^*e^{i(\omega_3-\omega_1)t}(|1\rangle\langle1|(t)-|3\rangle\langle3|(t))+i\Omega_2^*e^{i(\omega_3-\omega_2)t}|2\rangle\langle1|(t)\nonumber\\
                               &&+i(|1\rangle\langle1|(t)-|3\rangle\langle3|(t))(\xi_a(t)+\xi_a^{\dag}(t))+i|2\rangle\langle1|(t)(\xi_b(t)+\xi_b^{\dag}(t)),\nonumber\\
\frac{d}{dt}|3\rangle\langle2|(t)&=&i(\omega_3-\omega_2)|3\rangle\langle2|(t)+i\Omega_2^*e^{i(\omega_3-\omega_2)t}(|2\rangle\langle2|(t)-|3\rangle\langle3|(t))+i\Omega_1^*e^{i(\omega_3-\omega_1)t}|1\rangle\langle2|(t)\nonumber\\
                               &&+i(|2\rangle\langle2|(t)-|3\rangle\langle3|(t))(\xi_b(t)+\xi_b^{\dag}(t))+i|1\rangle\langle2|(t)(\xi_a(t)+\xi_a^{\dag}(t)),\nonumber\\
\frac{d}{dt}|2\rangle\langle1|(t)&=&i(\omega_2-\omega_1)|2\rangle\langle1|(t)+i\Omega_2e^{-i(\omega_3-\omega_2)}|3\rangle\langle1|(t)-i\Omega_1^*e^{i(\omega_3-\omega_1)t}|2\rangle\langle3|(t)\nonumber\\
                               &&+i|3\rangle\langle1|(t)(\xi_b(t)+\xi_b^{\dag}(t))-i|2\rangle\langle3|(t)(\xi_a(t)+\xi_a^{\dag}(t)),
\label{Heq}
\end{eqnarray}
with $|1\rangle\langle3|(t)$, $|2\rangle\langle3|(t)$, and $|2\rangle\langle1|(t)$ equal to the conjugates of
$|3\rangle\langle1|(t)$, $|3\rangle\langle2|(t)$, and $|1\rangle\langle2|(t)$, respectively.
Here $|i\rangle\langle{j}|(t)\equiv U^{\dag}(t)|i\rangle\langle{j}|U(t)$, and the noise operators $\xi_a(t)$ and $\xi_b(t)$ are defined as $\xi_a(t)\equiv\sum_kg_{a,k}e^{-i\omega_kt}a_k$ and $\xi_b(t)\equiv\sum_kg_{b,k}e^{-i\omega_kt}b_k$. Also, we have considered the resonant case with $\omega_{a}=\omega_3-\omega_1$ and $\omega_{b}=\omega_3-\omega_2$. Note that Eq.~(\ref{Heq}) is just the quantum Langvin equation of the system.

Let us introduce the Bargmann coherent states for the two bosonic baths,
\begin{equation}
|z\rangle\equiv \bigotimes_{k}| z_{k}\rangle=e^{\sum_{k}z_{ak}a_{k}^\dagger+\sum_{k}z_{bk}b_{k}^\dagger}|0\rangle,
\label{bhs}
\end{equation}
which satisfies
\begin{equation}
a_k|z\rangle=z_{ak}|z\rangle,~~~a_k^{\dag}|z\rangle=\frac{\partial}{\partial z_{ak}}|z\rangle,~~~b_k|z\rangle=z_{bk}|z\rangle,~~~b_k^{\dag}|z\rangle=\frac{\partial}{\partial z_{bk}}|z\rangle.
\end{equation}
As in \cite{Zhou} for the spin-boson model, we project Eq.~(\ref{Heq}) onto the Bargmann coherent states and take the expectation value for each operator. Then, we convert the quantum Langevin equation in Eq.~(\ref{Heq}) to
\begin{eqnarray}
\frac{\partial}{\partial t}\mathcal{A}(t,z)&=&\mathcal{H}(t)\mathcal{A}(t,z)+\left[z_{at}+\int_0^tds\alpha_a(t,s)\frac{\delta}{\delta z_{as}}\right]\mathcal{L}_a\mathcal{A}(t,z)+\left[z_{bt}+\int_0^tds\alpha_b(t,s)\frac{\delta}{\delta z_{bs}}\right]\mathcal{L}_b\mathcal{A}(t,z),
\label{sbe2}
\end{eqnarray}
where $\mathcal{A}(t,z)\equiv(\mathcal{A}_{11}(t,z),\mathcal{A}_{22}(t,z),\mathcal{A}_{33}(t,z),\mathcal{A}_{31}(t,z),
\mathcal{A}_{32}(t,z),\mathcal{A}_{21}(t,z),\mathcal{A}_{13}(t,z),\mathcal{A}_{23}(t,z),\mathcal{A}_{12}(t,z))^{T}$, with $T$ denoting the transpose of a matrix and $\mathcal{A}_{ij}(t,z)\equiv\langle\Psi_0|i\rangle\langle{j}|(t)|z\rangle\langle{z}|\Psi_0\rangle$. The matrix $\mathcal{H}(t)$ is given in Eq.~(\ref{matrixH}), and the matrices $\mathcal{L}_a$ and $\mathcal{L}_b$ are given in Eq.~(\ref{matrixL}). Also, in deriving Eq.~(\ref{sbe2}), we have used the functional chain rule
\begin{equation}
\frac{\partial}{\partial z_k}=\int ds\frac{\partial z_s}{\partial z_k}\frac{\delta}{\delta z_s},
\end{equation}
and defined the noise functions,
\begin{equation}
z_{at}=\sum_{k}g_{a,k}e^{-i\omega_kt}z_{ak}, ~~~z_{bt}=\sum_{k}g_{b,k}e^{-i\omega_kt}z_{bk}.
\end{equation}

Below we solve Eq.~(\ref{sbe2}) using a hierarchical-equation appraoch~\cite{Strunz3}. We extend the limits of integrals in Eq.~(\ref{sbe2}) to infinity and define $D_a\equiv\int_{-\infty}^{+\infty}\alpha_a(t,s)\frac{\delta}{\delta z_{as}}$ and $D_b\equiv\int_{-\infty}^{+\infty}\alpha_b(t,s)\frac{\delta}{\delta z_{bs}}$. According to Eq.~(\ref{sbe2}), $\mathcal{A}(t,z)$ only contains the noise for $0\leq s<t$. Then, for other values of $s$, we have $\frac{\delta}{\delta z_{is}}\mathcal{A}(t,z)=0$, where $i=a,b$. Thus, the extension of the integral limit is exact. Now Eq.~(\ref{sbe2}) can be expressed as
\begin{eqnarray}
\frac{\partial}{\partial t}\mathcal{A}(t,z)&=&\mathcal{H}(t)\mathcal{A}(t,z)+(z_{at}+D_a)\mathcal{L}_a\mathcal{A}(t,z)
+(z_{bt}+D_b)\mathcal{L}_b\mathcal{A}(t,z).
\label{sbe3}
\end{eqnarray}
We further define $\mathcal{A}^{(m,n)}(t,z)\equiv D_a^mD_b^n\mathcal{A}(t,z)$ and consider the Ornstein-Uhlenbeck correlation functions $\alpha_i(t,s)=\frac{\Gamma_i\gamma_i}{2}e^{-\gamma_i|t-s|}$, with $i=a,b$. Then, it follows that
\begin{eqnarray}
\frac{\partial}{\partial t}\mathcal{A}^{(m,n)}(t,z)&=&\frac{\partial}{\partial t}\left[\int_{-\infty}^{\infty}ds\alpha_a(t,s)\frac{\delta}{\delta z_{as}}\right]^m\left[\int_{-\infty}^{\infty}ds\alpha_b(t,s)\frac{\delta}{\delta z_{bs}}\right]^n\mathcal{A}(t,z)\nonumber\\
                        &=&-(m\gamma_a+n\gamma_b)\mathcal{A}^{(m,n)}(t,z)+\left[\int_{-\infty}^{\infty}ds\alpha_a(t,s)\frac{\delta}{\delta z_{as}}\right]^m\left[\int_{-\infty}^{\infty}ds\alpha_b(t,s)\frac{\delta}{\delta z_{bs}}\right]^n\frac{\partial}{\partial t}\mathcal{A}(t,z).\nonumber\\
                        &=&-(m\gamma_a+n\gamma_b)\mathcal{A}^{(m,n)}(t,z)+D_a^mD_b^n\frac{\partial}{\partial t}\mathcal{A}(t,z).
\label{eq-mn}
\end{eqnarray}
Using Eq.~(\ref{sbe3}), we obtain
\begin{eqnarray}
D_a^mD_b^n\frac{\partial}{\partial t}\mathcal{A}(t,z)&=&D_a^mD_b^n\left[\mathcal{H}(t)\mathcal{A}(t,z)+z_{at}\mathcal{L}_a\mathcal{A}(t,z)+\mathcal{L}_aD_{at}\mathcal{A}(t,z)+z_{bt}\mathcal{L}_b\mathcal{A}(t,z)+\mathcal{L}_bD_{bt}\mathcal{A}(t,z)\right]\nonumber\\
                  &=&D_{at}^mD_{bt}^n\mathcal{H}(t)\mathcal{A}(t,z)+D_{at}^mD_{bt}^nz_{at}\mathcal{L}_a\mathcal{A}(t,z)+D_{at}^mD_{bt}^n\mathcal{L}_aD_{at}\mathcal{A}(t,z)\nonumber\\
                                                 &&+D_{at}^mD_{bt}^nz_{bt}\mathcal{L}_b\mathcal{A}(t,z)+D_{at}^mD_{bt}^n\mathcal{L}_bD_{bt}\mathcal{A}(t,z)\nonumber\\
                                                 &=&\mathcal{H}(t)\mathcal{A}^{(m,n)}(t,z)+m\frac{\Gamma_a\gamma_a}{2}\mathcal{L}_aD_{at}^{m-1}D_{bt}^n\mathcal{A}(t,z)+z_{at}\mathcal{L}_a\mathcal{A}^{(m,n)}(t,z)\nonumber\\
                                                 &&+\mathcal{L}_aD_{at}^{m+1}D_{bt}^n\mathcal{A}(t,z)+n\frac{\Gamma_b\gamma_b}{2}\mathcal{L}_bD_{at}^mD_{bt}^{n-1}\mathcal{A}(t,z)+z_{bt}\mathcal{L}_bD_{at}^mD_{bt}^n\mathcal{A}(t,z)+\mathcal{L}_bD_{at}^mD_{bt}^{n+1}\mathcal{A}(t,z)\nonumber\\
                                                 &=&\mathcal{H}(t)\mathcal{A}^{(m,n)}(t,z)+m\frac{\Gamma_a\gamma_a}{2}\mathcal{L}_a\mathcal{A}^{(m-1,n)}(t,z)+z_{at}\mathcal{L}_a\mathcal{A}^{(m,n)}(t,z)\nonumber\\
                                                 &&+\mathcal{L}_a\mathcal{A}^{(m+1,n)}(t,z)+n\frac{\Gamma_b\gamma_b}{2}\mathcal{L}_b\mathcal{A}^{(m,n-1)}(t,z)+z_{bt}\mathcal{L}_b\mathcal{A}^{(m,n)}(t,z)+\mathcal{L}_b\mathcal{A}^{(m,n+1)}(t,z).
\label{eq-mn-1}
\end{eqnarray}
Substitution of Eq.~(\ref{eq-mn-1}) into Eq.~(\ref{eq-mn}) gives
\begin{eqnarray}
\frac{\partial}{\partial t}\mathcal{A}^{(m,n)}(t,z)
                        &=&-(m\gamma_a+n\gamma_b)\mathcal{A}^{(m,n)}(t,z)+\mathcal{H}(t)\mathcal{A}^{(m,n)}(t,z)+m\frac{\Gamma_a\gamma_a}{2}\mathcal{L}_a\mathcal{A}^{(m-1,n)}(t,z)+z_{at}\mathcal{L}_a\mathcal{A}^{(m,n)}(t,z)\nonumber\\
                                                 &&+\mathcal{L}_a\mathcal{A}^{(m+1,n)}(t,z)+n\frac{\Gamma_b\gamma_b}{2}\mathcal{L}_b\mathcal{A}^{(m,n-1)}(t,z)+z_{bt}\mathcal{L}_b\mathcal{A}^{(m,n)}(t,z)+\mathcal{L}_b\mathcal{A}^{(m,n+1)}(t,z).
\label{sbe4}
\end{eqnarray}
This is a stochastic differential equation, because it involves the noise functions $z_{at}$ and $z_{bt}$.

Let $\mathcal{A}^{(m,n)}(t)\equiv\mathcal{M}\{\mathcal{A}^{(m,n)}(t,z)\}$, where the statistical average is defined as
\begin{eqnarray}
\mathcal{M}\left\{\dots\right\}\equiv\prod_k\int\frac{d^2z_{ak}d^2z_{bk}}{\pi^2}e^{-|z_{ak}|^2-|z_{bk}|^2}\left\{\dots\right\}.
\label{oc}
\end{eqnarray}
It follows from Eq.~(\ref{sbe4}) that
\begin{eqnarray}
\frac{\partial}{\partial t}\mathcal{A}^{(m,n)}(t)&=&\mathcal{M}\left\{\frac{\partial}{\partial t}\mathcal{A}^{(m,n)}(t,z)\right\}\nonumber\\
                        &=&\mathcal{M}\left\{z_{at}\mathcal{L}_a\mathcal{A}^{(m,n)}(t,z)\right\}+\mathcal{M}\left\{z_{bt}\mathcal{L}_b\mathcal{A}^{(m,n)}(t,z)\right\}\nonumber\\
                         &&-(m\gamma_a+n\gamma_b)\mathcal{A}^{(m,n)}(t)+\mathcal{H}(t)\mathcal{A}^{(m,n)}(t)+m\frac{\Gamma_a\gamma_a}{2}\mathcal{L}_a\mathcal{A}^{(m-1,n)}(t)+\mathcal{L}_a\mathcal{A}^{(m+1,n)}(t)\nonumber\\
                         &&+n\frac{\Gamma_b\gamma_b}{2}\mathcal{L}_b\mathcal{A}^{(m,n-1)}(t)+\mathcal{L}_b\mathcal{A}^{(m,n+1)}(t).
\label{be1}
\end{eqnarray}

For an arbitrary element $\mathcal{A}^{(m,n)}_{ij}(t,z)$ of $\mathcal{A}^{(m,n)}(t,z)$,
\begin{eqnarray}
\mathcal{M}\left\{z_{at}\mathcal{A}^{(m,n)}_{ij}(t,z)\right\}&=&\mathcal{M}\left\{z_{at}D_{at}^mD_{bt}^n\langle\Psi_0|i\rangle\langle{j}|(t)|z\rangle\langle{z}|\Psi_0\rangle\right\}\nonumber\\
                                                              &=&\mathcal{M}\left\{\langle\Psi_0|i\rangle\langle{j}|(t)[\xi_a^\dag(t)]^m[\xi_b^\dag(t)]^n\xi_a(t)|z\rangle\langle{z}|\Psi_0\rangle\right\}\nonumber\\
                                                              &=&\langle\Psi_0|i\rangle\langle{j}|(t)[\xi_a^\dag(t)]^m[\xi_b^\dag(t)]^n\xi_a(t)|\Psi_0\rangle\nonumber\\
                                                              &=&0,\nonumber
\end{eqnarray}
because $\xi_a(t)|\Psi_0\rangle=\sum_kg_{a,k}e^{-i\omega_kt}a_k|D(0)\rangle\otimes|0\rangle
=\sum_kg_{a,k}e^{-i\omega_kt}a_k|D(0)\rangle\otimes|0\rangle_a\otimes|0\rangle_b=0$.
Thus, we have $\mathcal{M}\left\{z_{at}\mathcal{L}_a\mathcal{A}^{(m,n)}(t,z)\right\}=0$.
Similarly, because $\xi_b(t)|\Psi_0\rangle=\sum_kg_{b,k}e^{-i\omega_kt}b_k|D(0)\rangle\otimes|0\rangle=0$, we can also derive that $\mathcal{M}\left\{z_{bt}\mathcal{L}_b\mathcal{A}^{(m,n)}(t,z)\right\}=0$. Therefore, Eq.~(\ref{be1}) is simplified as
\begin{eqnarray}
\frac{\partial}{\partial t}\mathcal{A}^{(m,n)}(t)
                        &=&-(m\gamma_a+n\gamma_b)\mathcal{A}^{(m,n)}(t)+\mathcal{H}(t)\mathcal{A}^{(m,n)}(t)+m\frac{\Gamma_a\gamma_a}{2}\mathcal{L}_a\mathcal{A}^{(m-1,n)}(t)+\mathcal{L}_a\mathcal{A}^{(m+1,n)}(t) \nonumber\\
                        &&+n\frac{\Gamma_b\gamma_b}{2}\mathcal{L}_b\mathcal{A}^{(m,n-1)}(t)+\mathcal{L}_b\mathcal{A}^{(m,n+1)}(t),
\label{hi-eq}
\end{eqnarray}
where $\mathcal{A}^{(m,n)}(t)=0$ if $m$ or $n<0$. This is the hierarchical equation given in Eq.~(\ref{be2}).

Moreover, there is the relation that
\begin{eqnarray}
\mathcal{A}^{(m,n)}_{ij}(t)&=&\mathcal{M}\left\{\mathcal{A}^{(m,n)}_{ij}(t,z)\right\}\nonumber\\
&=&\mathcal{M}\left\{D_{at}^mD_{bt}^n\langle\Psi_0|i\rangle\langle{j}|(t)|z\rangle\langle{z}|\Psi_0\rangle\right\}\nonumber\\
&=&\mathcal{M}\left\{\langle\Psi_0|i\rangle\langle{j}|(t)[\xi_a^\dag(t)]^m [\xi_b^\dag(t)]^n|z\rangle\langle{z}|\Psi_0\rangle\right\}\nonumber\\
&=&\langle\Psi_0|i\rangle\langle{j}|(t)[\xi_a^\dag(t)]^m[\xi_b^\dag(t)]^n|\Psi_0\rangle\nonumber.                                                         \end{eqnarray}
Because $\xi_a(0)|\Psi_0\rangle=0$, and $\xi_b(0)|\Psi_0\rangle=0$, then
$\mathcal{A}^{(m,n)}_{ij}(0)=\langle\Psi_0|i\rangle\langle{j}|[\xi_a^\dag(0)]^m[\xi_b^\dag(0)]^n|\Psi_0\rangle=0$,
if $m$ or $n\ne 0$. Therefore, the initial condition of Eq.~(\ref{hi-eq}) is $\mathcal{A}^{(m,n)}(0)=0$ for $m$ or $n\ne 0$.

Note that
\begin{eqnarray}
\mathcal{A}^{(0,0)}(t)=\mathcal{M}\left\{\mathcal{A}(t,z)\right\}=\prod_k\int\frac{d^2z_{ak}d^2z_{bk}}{\pi^2}e^{-|z_{ak}|^2-|z_{bk}|^2}
\mathcal{A}(t,z)\equiv\mathcal{A}(t),
\end{eqnarray}
where $\mathcal{A}(t)\equiv(\mathcal{A}_{ij}(t))$, with $i,j=1,2,3$, and $\mathcal{A}_{ij}(t)=\langle\Psi_0|i\rangle\langle j|(t)|\Psi_0\rangle$.
From Eq.~(\ref{hi-eq}), we get
\begin{equation}
\frac{\partial}{\partial t}\mathcal{A}(t)=\mathcal{H}(t)\mathcal{A}(t)+\mathcal{L}_a\mathcal{A}^{(1,0)}(t)
+\mathcal{L}_b\mathcal{A}^{(0,1)}(t),
\end{equation}
which is the non-Markovian quantum Bloch equation in Eq.~(\ref{qbe}).

\section*{Acknowledgements}

This work is supported by the National Key Research and Development Program of China (Grant No.~2016YFA0301200), and the NSAF (Grant Nos.~U1330201 and U1530401). L.-A.W. is supported by Spanish MINECO/FEDER Grant FIS2015-69983-P, Basque	 Government Grant IT986-16 and UPV/EHU UFI 11/55.
T.Y. is supported by the DOD/AF/AFOSR Grant No.~FA9550-12-1-0001, and he thanks CSRC for hospitality during his visit.

\section*{Author contributions statement}

Z.Y.Z. and M.C. performed the derivations and numerical calculations under the guidance of J.Q.Y.. Also, L.A.W. and T.Y. participated in the discussions. All authors contributed to the interpretation of the work and the writing of the manuscript.

\section*{Additional information}

\textbf{Competing financial interests}: The authors declare no competing financial interests.

\end{document}